\documentclass[twocolumn,superscriptaddress,prb,amsmath,amssymb]{revtex4}
\usepackage{graphicx}% Include figure files
\usepackage{dcolumn}% Align table columns on decimal point
\usepackage{bm}% bold math

\begin{document}

%\preprint{APS/123-QED}

\title{Transport, thermal and magnetic properties of
RuSr$_2$(Gd$_{1.5}$Ce$_{0.5}$)Cu$_2$O$_{10-\delta}$, \\ a magnetic
superconductor
}% Force line breaks with \\

\author{D. G. Naugle}
\email{naugle@physics.tamu.edu} \affiliation{Physics Department,
Texas A\&M University, TX 77843, USA}
 %\altaffiliation[Also at ]{Texas 77843-4242, USA}

\author{K. D. D. Rathnayaka}
\affiliation{Physics Department, Texas A\&M University, TX 77843,
USA}

\author{V. B. Krasovitsky}
\affiliation{B. Verkin Institute for Low Temperature Physics and
Engineering, National Academy of Sciences, pr. Lenina 47, Kharkov
61103, Ukraine}

\author{B. I. Belevtsev}
\email{belevtsev@ilt.kharkov.ua} \affiliation{B. Verkin Institute
for Low Temperature Physics and Engineering, National Academy of
Sciences, pr. Lenina 47, Kharkov 61103, Ukraine}

\author{M. P. Anatska}
\affiliation{Physics Department, Texas A\&M University, TX 77843,
USA}

\author{G. Agnolet}
\affiliation{Physics Department, Texas A\&M University, TX 77843,
USA}

\author{I. Felner}
\affiliation{Racah Institute of Physics, The Hebrew University,
Jerusalem, 91904, Israel}


\begin{abstract}
Resistivity, thermoelectric power, heat capacity and magnetization
for samples of
RuSr$_2$(Gd$_{1.5}$Ce$_{0.5}$)Cu$_{2}$O$_{10-\delta}$ were
investigated in the temperature range 1.8-300 K with a magnetic
field up to 8 T. The resistive transitions to the superconducting
state are found to be determined by the inhomogeneous (granular)
structure, characterized by the intragranular, $T_{c0}$, and
intergranular, $T_{cg}$, transition temperatures. Heat capacity,
$C(T)$, shows a jump at the superconducting transition temperature
$T_{c0}\approx 37.5$~K. A Schottky-like anomaly is found in $C(T)$
below 20 K. This low temperature anomaly can be attributed to
splitting of the ground term $^{8}S_{7/2}$ of paramagnetic
Gd$^{3+}$ ions by internal and external magnetic fields.
\end{abstract}

%\pacs{Valid PACS appear here}
\maketitle

RuSr$_2$(Gd$_{1.5}$Ce$_{0.5}$)Cu$_{2}$O$_{10-\delta}$ belongs to
a known family of magnetic superconductors
\cite{felner,lorenz,awana,belevtsev}. Superconductivity is associated with
CuO$_2$ planes, while magnetic order is thought to be connected
with RuO$_2$ planes. The exact nature of the magnetic order in
this compound is still unknown, but it is conjectured that below
80-100 K weak-ferromagnetic order dominates. The paramagnetic
magnetic moments of Gd$^{3+}$ ions make a considerable
contribution to total magnetization as well as the heat capacity,
especially at low
temperature. A possible magnetic Ru-Gd interaction cannot be
excluded as well. Superconductivity in this family of compounds is
apparent below 50 K, where both superconducting and magnetic order
coexist.
\par
In this report, the transport, magnetic and thermal properties of
samples of RuSr$_2$(Gd$_{1.5}$Ce$_{0.5}$)Cu$_{2}$O$_{10-\delta}$
as prepared (by a solid-state reaction method) and annealed (12
hours at 845$^{\circ}$C) in pure oxygen at 30, 62, 78 atm, are
presented. The measurements were made with Quantum Design
devices (PPMS and SQUID magnetometer), and a homemade thermopower
measuring system.
\par
The samples behave as inhomogeneous (granular) superconductors.
This manifests itself to the greatest extent in resistive
properties, as can be seen, for example, in Fig. 1 for an annealed
(62 atm) sample. The rather broad and shouldered $R(T)$ curves in
the region of the superconducting transition are indicative of
inhomogeneity effects. The most obvious inhomogeneity source is
the granular structure, determined by the polycrystalline
structure (with a grain size of a few $\mu$m). Non-homogeneous
oxygen distribution can cause oxygen depletion of the
grain-boundary regions and, hence, weak electrical connectivity
between the grains, as is often the case in cuprates. Above the
superconducting transition, the rather high resistivity (about
$10^{-2}$~$\Omega$~cm) and the weak increase in resistance with
decreasing temperature support this suggestion. The onset
temperature of superconductivity, $T_{c}^{onset}$, is about 58 K
with a zero-resistance temperature about 25 K. Derivatives
$dR(T)/dT$ reveal two peaks (Fig. 2) in the region of the
superconducting transition, the positions of which can be
attributed to intragranular and intergranular superconducting
transitions at temperatures $T_{c0}$ and $T_{cg}$, respectively.
The intergranular superconductivity should be determined by
Josephson coupling between the grains. $T_{c0}$ and $T_{cg}$ are
equal to 37.5 K and 32.8 K, respectively, in zero field. At the
maximum field 8 T used in this study, they reduce to 34.7 K and
12.4 K, respectively. Thus, the magnetic field has a weak
influence on the intragranular transition temperature $T_{c0}$.
The intergranular $T_{cg}$ is far more sensitive to magnetic
field, with the main variations occurring in the low field region
$H < 0.5$~T (Fig. 2).
\par
The $R(T)$ behavior of the as-prepared sample (Fig. 3 inset), 
which is expected
to be the most depleted in oxygen, substantiates our assumptions.
The $R(T)$ curve of this sample, taken at $H = 0$, indicates that
$T_{c0}$ and $T_{cg}$ are equal to 35 K and 18.5 K, respectively,
with $T_{c}^{onset}$ about 39 K. It is seen that intragranular
superconducting properties are much less affected by high-pressure
oxygen annealing than the intergranular ones.
\par
The thermoelectric power ($S$) of the samples studied is found to
be sensitive to oxygen annealing as well (Fig.~3). The magnitude
of $S$ at $T=290$~K is far (1.8 times) larger in the as prepared
sample than that in the sample annealed at 62 atm of O$_2$. The
temperatures of maximum slope in the $S(T)$ curves at the
superconducting transition (which can be taken as $T_c$ values)
are 41.5~K and 22.8 K in 62-atm annealed and as prepared samples,
respectively. Derivatives $dS/dT$ give estimated
values of $T_{c}^{onset}$ as well, which are found to be roughly
 44 K and 59~K for as-prepared and 62-atm annealed samples,
respectively, that correspond roughly to values obtained from the
resistive data (inset).
\par
The measured temperature dependences of the magnetization, $M(T)$,
indicate that $M$ becomes appreciable only below
180--200 K with a further sharp jump in $M$ in the range of 100 K at
$T=T_{m2}$ due to transition to a weak-ferromagnetic
state. The transition temperature $T_{m2}$ is about 99 K and 90 K
for as-prepared and 100-atm annealed samples. This shows that oxygen
annealing suppresses somewhat the magnetic order in
ruthenocuprates in agreement with previous studies
\cite{felner,awana}. $M(T)$ dependences (ZFC) recorded in low
field (about 0.5 mT) show a clear diamagnetic transition below 40
K determined by transition to the superconducting state. The values of
$T_c$ estimated from the diamagnetic part of $M(T)$ correspond
well to $T_{c0}$ obtained from transport properties.

\begin{figure}[t]
\includegraphics[width=0.8\linewidth]{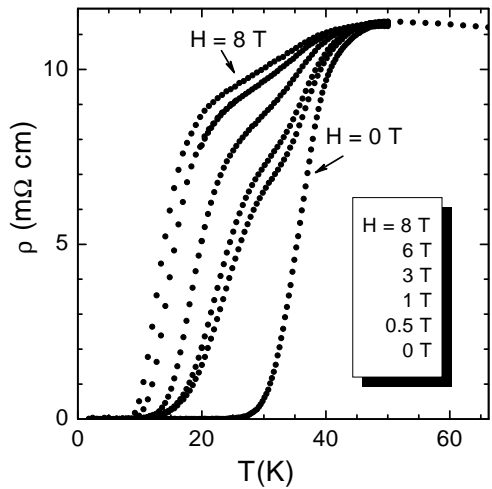}% Here is how to import EPS art
\caption{Temperature dependence of the resistivity $\rho(T)$ at
different magnetic fields for sample annealed in pure oxygen at
62 atm. }
\end{figure}

\begin{figure}[tb]
\includegraphics[width=0.8\linewidth]{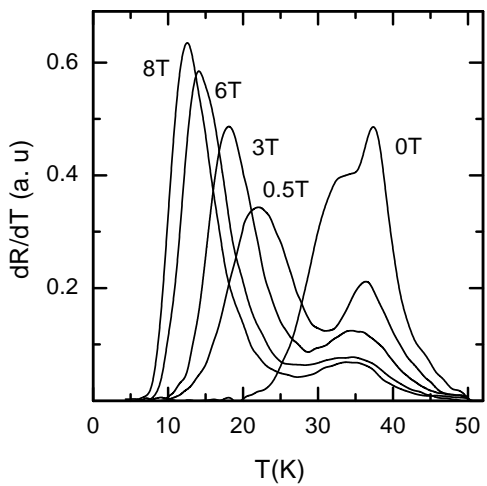}% Here is how to import EPS art
\caption{Derivatives of resistive curves of superconducting
transition (for the sample shown in Fig. 1) for different magnetic
fields.}
\end{figure}

\par
In contrast to transport properties, the heat capacity was found
to be rather insensitive to the granular structure and oxygen
annealing. The specific heat data are found to be nearly the same
for all 4 samples studied. No features in the temperature dependence
of the heat capacity, $C(T)$, associated with magnetic transitions
were found. This can be attributed to absence of long-range
magnetic order at the transition point due to magnetic
inhomogeneities induced by structural and/or stoichiometric
inhomogeneity, or phase separation. The low temperature behavior
of $C(T)$ (below 40~K) indicates the superconducting transition
and a magnetic anomaly. To present these features of $C(T)$ more
clearly, an estimate of the lattice heat capacity contribution,
$C_{ph}(T)$, was subtracted. Measurements of $C(T)$ of a
RuSr$_2$(Eu$_{1.5}$Ce$_{0.5}$)Cu$_{2}$O$_{10-\delta}$ sample of
the same composition but with non-magnetic Eu substituted for
magnetic Gd were used for this purpose. The Eu sample displayed a
smooth $C(T)$ dependence of the Debye type without any low
temperature magnetic anomaly or jump at the superconducting
transition. The function $\Delta C(T)\approx C_{Gd}- bC_{Eu}$
($C_{Gd}$ is as measured data for Gd sample, $C_{Eu}$ is that for
Eu-sample, $b$ is a factor close to unity, providing $C_{Gd}=
bC_{Eu}$ at $T=300$~K) is shown in Fig. 4.
\par
 The main features in $\Delta C(T)$ are (i) the jump at
the superconducting transition, and (ii) the upturn below 20~K
(Schottky-like anomaly). A jump at $T_c$ in the heat capacity of
a ruthenocuprate with a similar chemical composition
[RuSr$_2$(Gd$_{1.4}$Ce$_{0.6}$)Cu$_2$O$_{10-\delta}$] was reported
earlier \cite{chen}, but no Schottky-like anomaly was previously
reported.

\begin{figure}[t]
\includegraphics[width=0.85\linewidth]{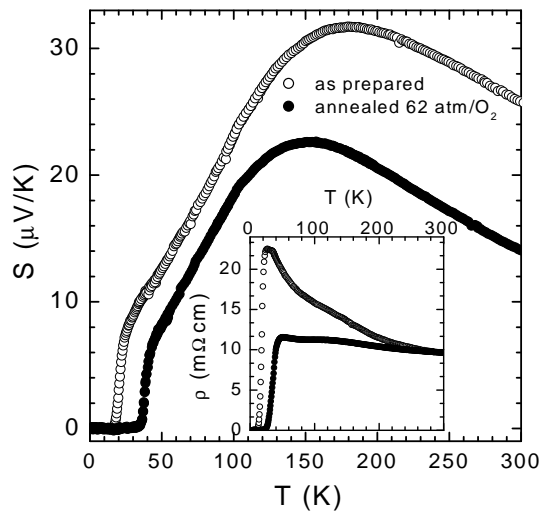}% Here is how to import EPS art
\vspace{-6pt} \caption{Temperature dependences of the
thermoelectric power for as prepared and 62 atm/O$_2$ annealed
samples. The inset shows $\rho(T)$ at zero field for the same
samples.
 }
\end{figure}

\par
In heat capacity studies, the temperature of the superconducting
transition is usually associated with onset of the $C(T)$ jump
($T_{c}^{o}$) or with the point of maximum slope of $C(T)$ in this
region. In zero field $T_{c}^{o} \approx 37$~K which is very close
to the intragranular $T_{c0}$ determined from the $R(T)$ curves.
An external field up to $H = 8$~T hardly produces shifts in
$T_{c}^{o}$ (Fig. 4). This correlates well with the very weak
shift in $T_{c0}$ in a magnetic field (Fig. 2) and provides evidence of
enormous upper critical fields in the ruthenocuprates.
\par
The low temperature Schottky-like anomaly can be attributed to
splitting of the ground term $^{8}S_{7/2}$ of paramagnetic
Gd$^{3+}$ ions. According to Kramers' theorem, the degenerate
ground term should be split into 4 doublets in tetragonal
symmetry. The sources of splitting can be crystal-electric-field
effects and the internal and external magnetic fields. The
crystal-field effect can be ignored, in the first approximation,
since Gd$^{3+}$ has zero orbital angular momentum. Other sources
of splitting cannot be, however, excluded. In particular, internal
molecular fields can arise in the ruthenocuprate from both the Gd
and Ru sublattices \cite{tallon} and can coexist with
superconductivity \cite{leviev}. Even though a direct Gd-Gd exchange
interaction is unlikely, these ions can be magnetically polarized
by the $4d$-$4f$ interaction.

\begin{figure}[t]
\includegraphics[width=0.85\linewidth]{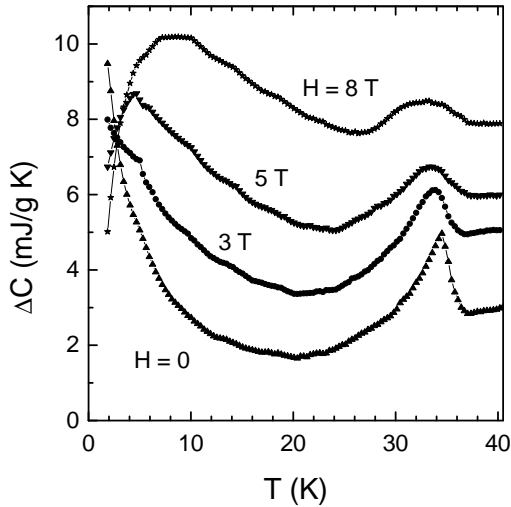}% Here is how to import EPS art
\vspace{-6pt} \caption{Temperature curves of the non-phonon part
of the heat capacity, measured in different magnetic fields.
 }
\end{figure}

\par
Generally, the Schottky term, $C_{Sch}(T)$, in the heat capacity
for compounds with Gd$^{3+}$ ions should be attributed to
splitting of all 4 doublets, although actually only some of them
make the dominant contribution to the effect. For any number of
influencing levels, however, the function $C_{Sch}(T)$ should
have a maximum at a temperature $T_{max}$, which is of the order
of $\Delta_{s}/k_{B}$. Here the $\Delta_{s}$ is a characteristic
energy-level splitting, which is equal to  $2\mu_{eff}(H_{mf} +
H)$, where $\mu_{eff}$ is the effective moment of Gd ions,
$H_{mf}$ is the effective molecular field at the Gd$^{3+}$ site
and $H$ is the external field. The field $H_{mf}$ can depend on
external field as well. It is clear that $T_{max}$ should
increase and the maximum should become more smeared with
increasing external field, which is the case (Fig.~4). Thus,
the observed low-temperature heat capacity anomaly is in all
probability connected with the Schottky effect. An exact
numerical analysis of the effect in the sample studied presents a
real challenge since most of the important parameters (influence
of magnetic field on schemes of energy levels, molecular field,
effective moments of Gd$^{3+}$ ions) are not known very well.

\par
This research was supported by the Robert A. Welch Foundation
(A-0514, A-1386), NSF (DMR-0103455, DMR-0315478, DMR-0422949) and
CRDF (UPI-2566-KH-03).

\end{document}